\documentclass[fleqn,usenatbib]{mnras}
\usepackage{longtable}
\usepackage{newtxtext,newtxmath}
\usepackage{lineno}
\usepackage{subfigure}
\usepackage[T1]{fontenc}
\usepackage{booktabs}
\usepackage{float}
\usepackage{makecell}
\usepackage{caption}
\DeclareRobustCommand{\VAN}[3]{#2}
\let\VANthebibliography\thebibliography
\def\thebibliography{\DeclareRobustCommand{\VAN}[3]{##3}\VANthebibliography}

\usepackage{graphicx}	
\usepackage{amsmath}	

\title[Metallicities of CCSNe]{A statistical study of the metallicity of core-collapse supernovae based on VLT/MUSE integral-field-unit spectroscopy}

\author[Xi et al.]{
Qiang Xi$^{1,2}$,
Ning-Chen Sun$^{1,2,3}$\thanks{E-mail: sunnc@ucas.ac.cn},
Yi-Han Zhao$^{1,2}$,
Justyn R. Maund$^{4}$,
Zexi Niu$^{1,2}$,
\newauthor
Adam J. Singleton$^{5}$,
Jifeng Liu$^{1,2,3,6}$
\\ 
$^{1}$School of Astronomy and Space Science, University of Chinese Academy of Sciences, Beijing 100049, People’s Republic of China\\
$^{2}$National Astronomical Observatories, Chinese Academy of Sciences, Beijing 100101, People’s Republic of China\\
$^{3}$Institute for Frontiers in Astronomy and Astrophysics, Beijing Normal University, Beijing, 102206, People’s Republic of China\\
$^{4}$Department of Physics, Royal Holloway, University of London, Egham, TW20 0EX, United Kingdom\\
$^{5}$School of Mathematical and Physical Sciences, University of Sheffield, Hicks Building, Hounsfield Road, Sheffield S3 7RH, United Kingdom\\
$^{6}$New Cornerstone Science Laboratory, National Astronomical Observatories, Chinese Academy of Sciences, Beijing 100012, People’s Republic of China
}

\date{Accepted XXX. Received YYY; in original form ZZZ}

\pubyear{\the\year{}}

\begin{document}
\label{firstpage}
\pagerange{\pageref{firstpage}--\pageref{lastpage}}
\maketitle

\begin{abstract}
Metallicity plays a crucial role in the evolution of massive stars and their final core-collapse supernova (CCSN) explosions. Integral-field-unit (IFU) spectroscopy can provide a spatially resolved view of SN host galaxies and serve as a powerful tool to study SN metallicities. While early transient surveys targeted on high star formation rate and metallicity galaxies, recent untargeted, wide-field surveys (e.g., ASAS-SN, ZTF) have discovered large numbers of SNe without this bias. In this work, we construct a large sample of SNe discovered by wide-field untargted searches, consisting of 166 SNe of Types II(P), IIn, IIb, Ib and Ic at $z \leq 0.02$ with VLT/MUSE observations. This is currently the largest CCSN sample with IFU observations. With the strong-line method, we reveal the spatially-resolved metallicity maps of the SN host galaxies and acquire accurate metallicity measurements for the SN sites, finding a range from $12 + \log(\text{O/H}) = 8.1$ to 8.7~dex. And the metallicity distributions for different SN types are very close to each other, with mean and median values of 8.4--8.5~dex. Our large sample size narrows the 1$\sigma$ uncertainty down to only 0.05~dex. The apparent metallicity differences among SN types are all within $\sim$1$\sigma$ uncertainties and the metallicity distributions for different SN types are all consistent with being randomly drawn from the same reference distribution. This suggests that metallicity plays a minor role in the origin of different CCSN types and some other metallicity-insensitive processes, such as binary interaction, dominate the distinction of CCSN types.
\end{abstract}




\section{Introduction}

Supernovae (SNe) are one of the most energetic catastrophic events in the Universe. They are categorized into Type I and Type II based on the presence of hydrogen lines in their spectra \citep{Minkowski1941}. Other than the thermonuclear Type Ia SNe, the other types originate from the core collapse (CC) of massive stars with initial masses of $\gtrsim$8~$M_{\sun}$ \citep{bethe1979equation,woosley1986physics,arnett1989supernova}. Most hydrogen-rich SNe are of Type~IIP, characterised by a plateau phase, powered by hydrogen recombination, in the light curve \citep{Barbon1979}. A fraction of SNe, classified as Type IIn, exhibit narrow emission lines in their spectra arising from the strong interaction between the fast SN ejecta and slow circumstellar material (CSM; \citealp{Schlegel1990}). Type~Ib and Type~Ic SNe are characterized by the absence of hydrogen features in their spectra, with Type Ic SNe also lacking helium features \citep{OPTICALSPECTRA}. As an intermediate class between the hydrogen-rich and hydrogen-poor SNe, Type IIb displays hydrogen lines in the early phases of the explosion, resembling Type II, but these features disappear quickly in the later stages,appearing similar to Type Ib \citep{nomoto1993type}. For Types IIb, Ib and Ic, the disappearance or lack of hydrogen/helium features are due to the stripping of the outer envelopes of their progenitor stars. They are also known, therefore, as the stripped-envelope (SE) SNe.

It is a major goal, and currently a major difficulty, to determine the progenitor stars of different types of SNe. Current research suggests that the progenitors of Type IIP SNe are red supergiants (RSGs) with massive hydrogen envelopes \citep{smartt2009}. However, stellar evolutionary theories are inconsistent with the lack of detection of high-mass ($>$16--18~M$_\odot$) RSG progenitors (i.e. the "RSG problem"; \citealp{smartt2009}).
This could result from CSM extinction underestimating the progenitor RSG’s mass \citep{Walmswell2012,Beasor2024} or from high-mass RSGs collapsing directly into black holes without a SN \citep{Kochanek2014}.It is also unclear to what extent binary interactions dominate versus contribute to the observed transient diversity\citep{Zapartas2021, Bostroem2023}.

 While luminous blue variables (LBVs) have been proposed as Type~IIn SN progenitors \citep{Gal-Yam2007,Kiewe2012,Smith2014araa,Elias-Rosa2016,Niu20242010jl}, it is still unclear why these stars undergo intense outbursts, creating the dense CSM, shortly before explosion. SESNe could originate from single massive WR stars \citep{woosley1986physics}, stripped by wind, or binary systems, where the progenitor is stripped by a companion star \citep{Podsiadlowski1992,Crockett2008,Folatelli_2015,Maeda2006,Maeda2014,Maeda2015,Fang_2018,Lyman2016,Taddia2018,Woosley_2021,Niu202417gkk,Zhao2025}. It still remains an open question what fraction of SESNe each channel contributes to.

For massive stars, stellar mass is the most important parameter that determines their structure and evolution. In addition, metallicity also play crucial roles; at high metallicties, stars have stronger line-driven winds, allowing for the stripping of the envelope and the formation of CSM \citep{Castor1975}. These effects can determine the light curve and spectral features, and even the classification, of their final SN explosion. Environmental studies offer a powerful approach to investigate the metallicity of CCSNe. During the short lifetimes ($\lesssim$50 million years) of massive stars, they can travel only a short distance from the formation to explosion sites and the environment has limited chemical evolution over such short timescales \citep{Anderson_James_Habergham_Galbany_Kuncarayakti_2015}.

Early studies on SN metallicity relied on long-slit spectroscopy \citep{Anderson2010,Modjaz2011, Leloudas2011,Sanders2012,Taddia2015} or even used the metallicity of the entire host galaxy as a proxy \citep{Langer2006,Prieto2008,Neill2011,Lunnan2014}. It has been suggested, however, that a high spatial resolution is necessary for the accurate measurement of SN metallicity based on gas emission lines from the environment \citep{Niino2015}. In more recent years, integral-field-unit (IFU) spectroscopy has been used to investigate SN metallicity \citep{Kuncarayakti2012Mass,Kuncarayakti2013,Kuncarayakti2013b,Kuncarayakti2015,Kuncarayakti2018,Galbany2014,galbany2016,Galbany2018,pessi2023,Moriya2023}. Instead of a single point or slit, IFU has the capability of acquiring spatially resolved spectral information over a relatively large fields of view.
This is important to reveal the complexity of the SN environment. A generally increasing trend in metallicity has been suggested  for IIP $\rightarrow$ IIb $\rightarrow$ Ib $\rightarrow$ Ic, correlated with the degree of envelope stripping.

\begin{figure}
    \centering
    \includegraphics[width=1\linewidth]{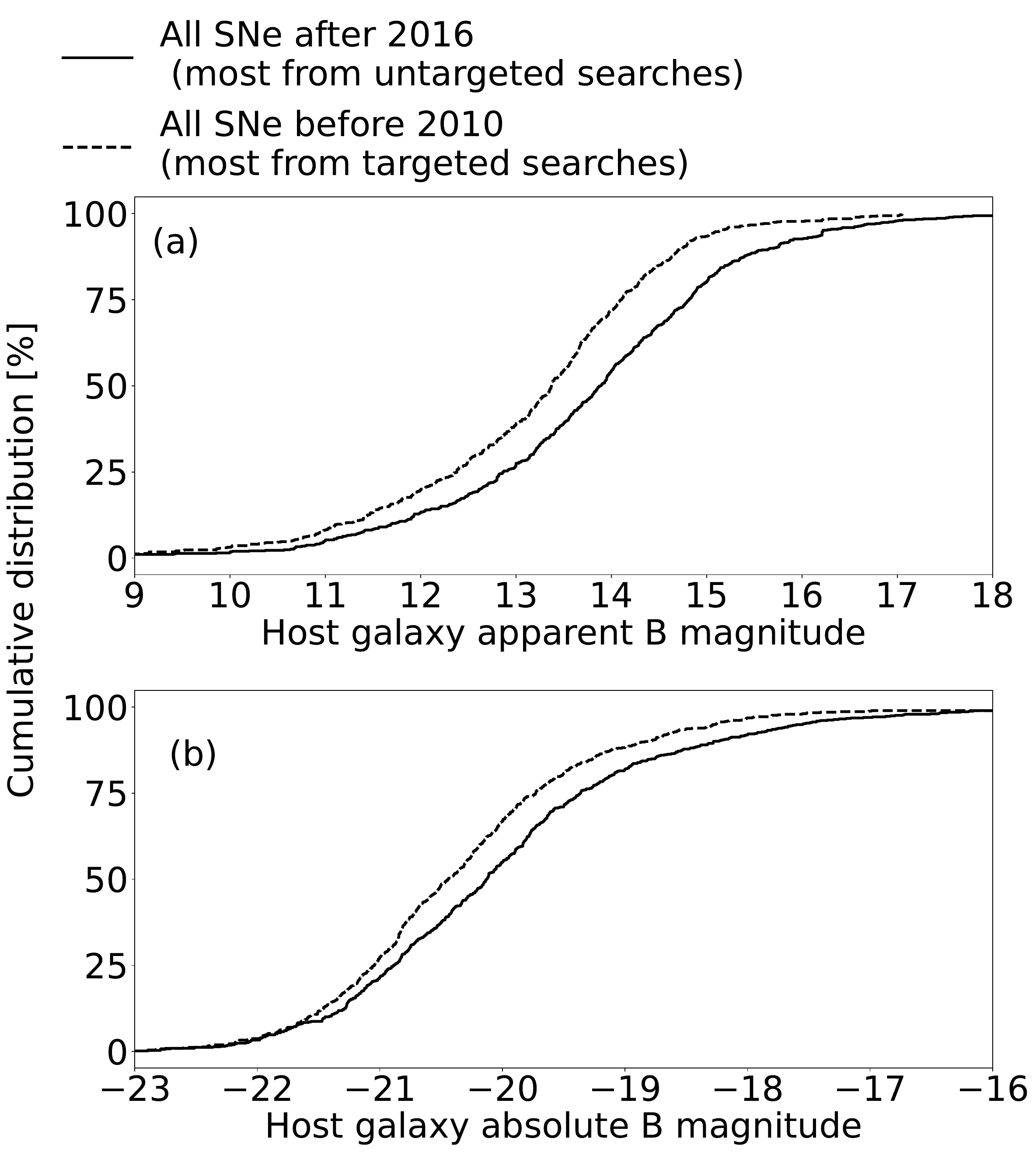}
    \caption{Cumulative distributions of the apparent (a) and absolute (b) B-band magnitudes of SN host galaxies. The dashed line is for SNe before 2010, when most were discovered by transient surveys targeted on bright galaxies, while the solid line is for those after 2016, when most were discovered by untargeted SN searches.}
    \label{fig:1}
\end{figure}

Within the domain of statistical research, minimizing sample bias is of particular importance. Limited by the telescopes' small field of view, early SN searches targeted on galaxies of high masses and star formation rates (SFRs) in order to maximize the number of discovered SNe. Such galaxies , however, also tend to have higher metallicities \citep{Tremonti_2004}, thus introducing a bias to SN samples discovered in this way \citep{Sanders2012}. With the increasing power of time-domain observations, more recent wide-field SN searches are able to map a significant portion of sky (e.g. the All Sky Automated Survey for SNe (ASAS-SN; \citealp{asassnKochanek_2017}). SNe discovered by such untargeted searches are not affected by the metallcity bias introduced by their host galaxies. Figure~\ref{fig:1} compares the host galaxy magnitudes of CCSNe discovered before 2010, when most were discovered by targeted searches, and after 2016, when most were discovered by untargeted searches. The SNe in this statistics are sourced from the Transient Name Server\footnote{\url{https://www.wis-tns.org/}} (TNS) and Open SN Catalog (OSC; \citealp{OSC2017}), with an additional selection criterion of redshift less than 0.05. At such proximities, the searches for SNe with typical luminosities are very complete. Magnitudes for SN host galaxies are from the GLADE+ catalog \citep{GELADE}. Some of the SNe lack this information for their host galaxies, and we have excluded these from the analysis. It is clear that SNe from targeted searches are significantly biased toward brighter host galaxies. Therefore, the early studies on SN metallicity are unavoidably affected by the bias caused by targeted SN discovery.

For studies that rely on archival observations, another potential bias may come from data availability as the archival observations are from different programs with different scientific goals, target selection criteria, observational strategies and even telescopes. Without a further careful selection, the sample could be rather heterogeneous with significant biases that are difficult to assess.


A large sample size is also very important to reduce the stochastic sampling effect. In this work, we study SN metallicity based on IFU observations carried out by the Multi-Unit Spectroscopic Explorer (MUSE) on the Very Large Telescope (VLT). With a careful selection, we build a large and untargeted sample of 166 CCSNe, which is to date the largest SN sample with IFU data. We try to look for metallicity difference among the SN types. Our aim is to explore the possible roles played by metallicity in the origin of CCSNe.

This paper is organized as follows: Section~\ref{method} explains our sample selection and metallicity measurement. Section~\ref{result} presents our results along with a discussion of key implications. Finally, this work is summarized in Section~\ref{summary}.

\section{Method} \label{method}

\subsection{Sample selection}

MUSE is an IFU instrument installed on the VLT operated by the European Southern Observatory (ESO) in Chile. It has a large field of view of $1\times 1$ arcmin$^2$ and covers a wavelength range from 4650 to 9300 \AA\ \citep{bacon2010muse}. This range covers the important gas emission lines (such as H$\alpha$, H$\beta$, [O~\textsc{iii}] $\lambda\lambda$4959, 5007,  and $[\mathrm{N~\textsc{ii}}]$ $\lambda\lambda$6548, 6583), with which metallicity can be derived with the strong-line method \citep{Pagel1979,Edmunds1984}. MUSE is, therefore, very suitable for SN metallicity studies.

As mentioned in the Introduction, further careful selection is curial to construct a minimally biased sample for the statistical analysis of SN metallicities. The two key considerations are SN discovery and data availability.

\paragraph*{SN discovery}

As mentioned in the Introduction, it is very important to avoid the bias by targeted SN searches. Therefore, we include in our sample only SNe discovered by the untargeted, wide-field transient surveys. Such surveys include the (Intermediate) Palomar Transient Factory (PTF; \citealp{Law2009ptf,Cao2016iptf}), ZTF \citep{Bellm_2019}, ASAS-SN \citep{asassnKochanek_2017}, Pan-STARRS \citep{chambers2019panstarrs1surveys}, the Asteroid Terrestrial-impact Last Alert System  (ATLAS; \citealp{Jedicke2012ALTAS})
, the Mobile Astronomical Systems of the Telescope-Robots (MASTER; \citealp{Lipunov2003MASTER}), Gaia \citep{Altavilla2012gaia}, the Catalina Sky Survey (CSS; \citealp{Christensen2014CSS}), the Sloan Digital Sky Survey (SDSS; \citealp{Frieman2008SDSS}) and La Silla-QUEST Variability Survey (LSQ; \citealp{Hadjiyska2012LSQ}). In addition, we require the redshift to be z $<$ 0.05, within which the giant HII regions of hundreds parsecs can be spatially resolved. We cross-match these SNe, queried from the TNS and OSC, with the ESO Data Archive\footnote{\url{https://archive.eso.org/scienceportal/home}}, getting 260 SNe with MUSE data.

\begin{figure}
    \centering
    \includegraphics[width=1\linewidth]{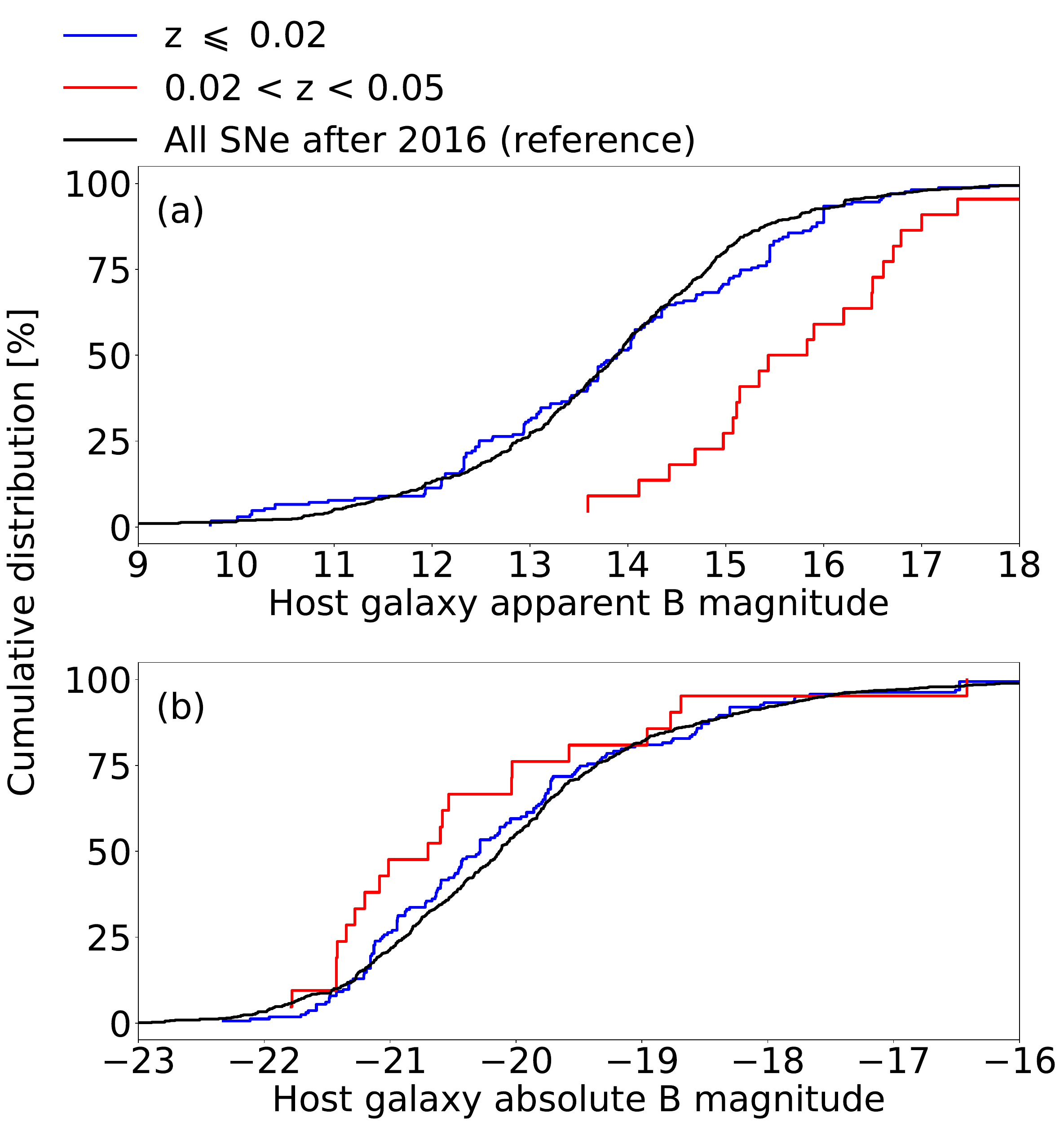}
    \caption{Cumulative distributions of the apparent (a) and absolute (b) B-band magnitudes of SN host galaxies. The black line is for all SNe after 2016, when most were discovered by untargeted SN searches (same as the black solid line in Fig.~\ref{fig:1}). They suffer little from discovery bias and are used as a reference distribution. The blue and red lines are SNe discovered by untargeted searches and with MUSE observations at redshift of $z \leq 0.02$ and $0.02 < z < 0.05$, respectively.}
    \label{fig:2}
\end{figure}

\paragraph*{Data availability}

Given that the wide field of view of the MUSE IFU spectrograph, the distant and low-mass galaxies with small angular diameters are less likely to be observed. Figure~\ref{fig:2} shows the host magnitude distribution for the above-selected SNe with MUSE data at different redshifts. For comparison, we use all SNe discovered after 2016 as a reference. It is clear that SNe with 0.02 $< z <$ 0.05 are systematically biased toward brighter host galaxies. On the other hand, SNe with $z \leq 0.02$ are similar to the reference sample. Therefore, we further apply a redshift cut of $z \leq 0.02$. Note that the MUSE data were compiled from different observing programs with different original scientific goals and target selection strategies. It is difficult to analyze the possible bias introduced by this heterogeneity. As shown in Figure~\ref{fig:2}, however, the sample at $z \leq 0.02$ is quite representative of the local SN population and we deem the possible bias could be small. It is also worth noting that, although each transient survey has its own limiting magnitude, cadence, and filter set, the CCSN subtypes considered in our sample — Type II(P), IIb, Ib, and Ic — exhibit very similar peak magnitudes and characteristic light‐curve timescales of order months. By restricting our sample to $z \leq 0.02$, we are confident that the discovery of local CCSNe by the current wide‐field surveys is complete out to this redshift.

\begin{figure}
    \centering
    \includegraphics[width=1.0\linewidth]{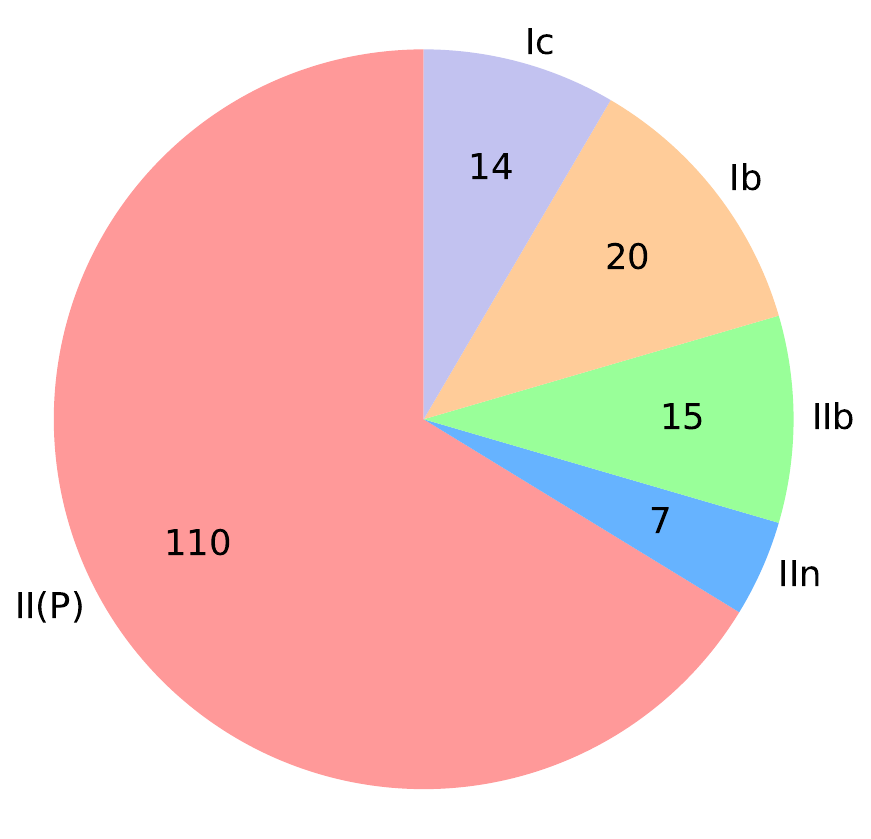}
    \caption{Number of SNe of different types in the final sample.}
    \label{fig:pie}
\end{figure}

After the above selection, there are 24 SNe of Type IIP, 7 of Type IIn, 14 of Type IIb, 20 SNe of Type Ib, 14 of Type Ic. In addition, there are 86 Type II SNe in our sample, for which the subtypes are unknown. Given that the overwhelming majority of Type II SNe are standard Type IIP SNe, we combined all SNe of Type IIP and Type II into a single subsample for analysis. We hereafter designate this subsample as Type II(P). Moreover, there are also 2 peculiar Type II, 3 peculiar Type Ib, 1 peculiar Type Ic, 1 Type Ibn, 1 Ca-rich Type Ib/Ic, 5 broad-lined Type Ic (Ic-BL), 1 Type Ib/c-BL, and 3 ambiguous Type IIn/LBV; these peculiar or ambiguous SNe are not in included in our analysis, leaving 166 SNe in the final sample. Details of the final sample are provided in Table~\ref{table_snlist}, and the distribution of SN types is shown in Figure~\ref{fig:pie}.

\subsection{Metallicity measurement}

The reduced MUSE datacube were obtained from the ESO Data Archive. We used the \textsc{ifuanal} package \citep{Lyman2018} to analyze the datacube. First we dereddened the datacube with Galactic extinction from \cite{Schlafly_2011}, and a standard extinction law with $R_V = 3.1$ \citep{1989ApJ...345..245C}. We then applied redshift corrections to the datacube with redshifts from OSC and  TNS. To acquire the spatially-resolved metallicity distribution across the galaxies, we employed the Voronoi binning with a target signal-to-noise ratio (SNR) of 120 within the wavelength range of 6540-6580 \AA, within which the H$\alpha + [\mathrm{N}~\textsc{ii}]$ lines reside. As will be described later, a minimum of 10 bins is required to fit the metallicity gradients; should fewer than 10 bins be found, we reduced the target SNR until 10 bins were achieved from the Voronoi binning. Due to differences in observation conditions and intrinsic galaxy properties, the number of bins for each galaxy varied from tens to several hundreds.

Inside each bin, we used \textsc{starlight} \citep{Fernandes2005starlight} to fit and remove the stellar continuum, leaving only the nebular emission lines from ionized gas. Gaussian fitting is used to derive the fluxes of lines including H$\alpha$, H$\beta$, [O~\textsc{iii}] $\lambda\lambda$ 4959, 5007,  and $[\mathrm{N~\textsc{ii}}]\lambda\lambda$6548, 6583. We determined the gas-phase metallicity using the strong-line method based on the O3N2 calibration from \cite{Marino2013A&A...559A.114M}. This method uses the ratio of strong lines with similar wavelengths, making it insensitive to extinction.
\begin{equation}
12 + \mathrm{log}(\mathrm{O/H}) = 8.533 -0.214\times \mathrm{O3N2},
\end{equation}
where
\begin{equation}
\mathrm{O3N2} = \log \left(\frac{[\mathrm{O~\textsc{iii}}] \lambda 5007}{\mathrm{H\beta}} \times \frac{\mathrm{H\alpha}}{[\mathrm{N~\textsc{ii}}] \lambda 6583}\right)
\end{equation}
For bins where [O~\textsc{iii}] or H$\beta$ were not detected (i.e., with amplitudes less than three times the spectral noise fluctuations), we used the N2 calibration instead
\begin{equation}
12 + \mathrm{log}(\mathrm{O/H}) = 8.743 +0.462\times \mathrm{N2},
\end{equation}
where
\begin{equation}
    \mathrm{N2} = \log \left(\frac{[\mathrm{N~\textsc{ii}}] \lambda 6583}{\mathrm{H\alpha}}\right)
\end{equation}
If the [N~\textsc{ii}] emission line was also too weak to be reliably detected, we tried to estimate an upper limit for the metallicity. Specifically, we derived the [N~\textsc{ii}] line width using the observed H$\alpha$ line width and the wavelength-dependent line spread function model of MUSE \citep{2guerou2017MUSE}. This width, combined with the 3$\sigma$ amplitude limit, allowed us to estimate an upper limit of the [N~\textsc{ii}] line flux, and in turn, an upper limit of the metallicity.

\begin{figure*}
    \centering
    \includegraphics[width=1.0\linewidth]{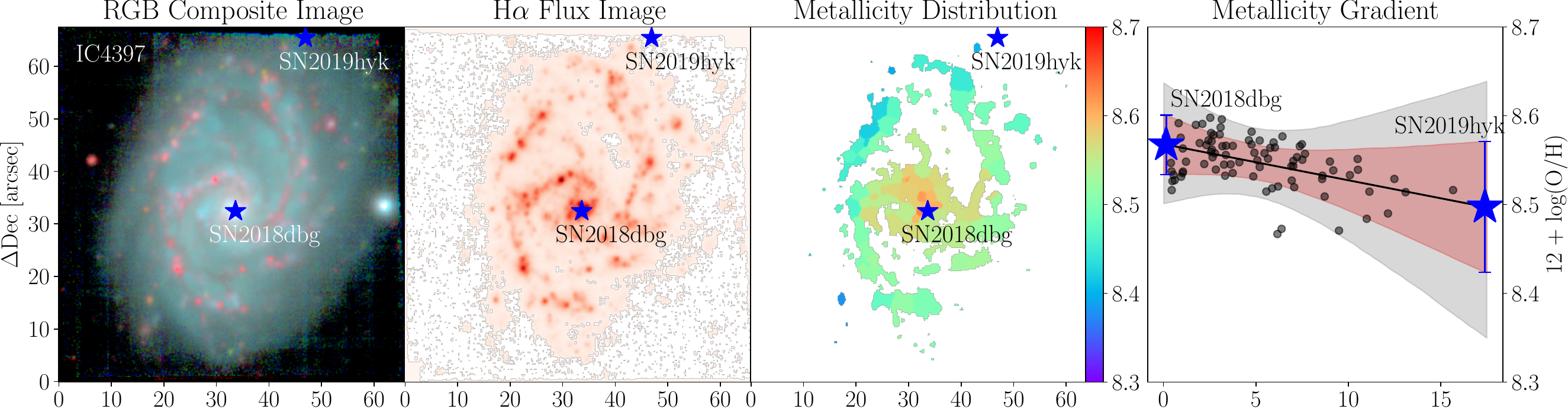}
    \includegraphics[width=1.0\linewidth]{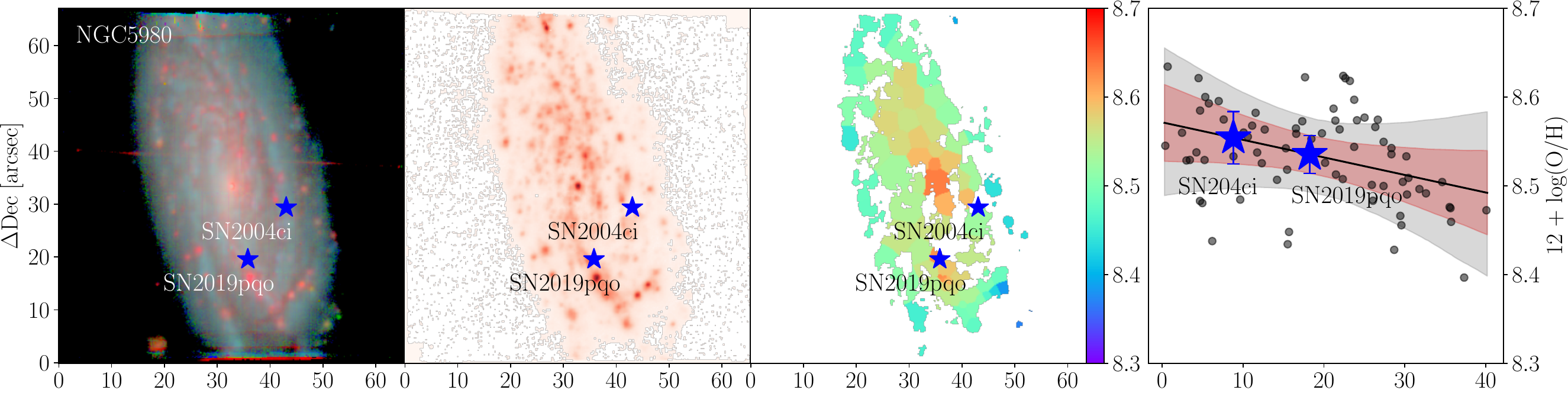}
    \includegraphics[width=1.0\linewidth]{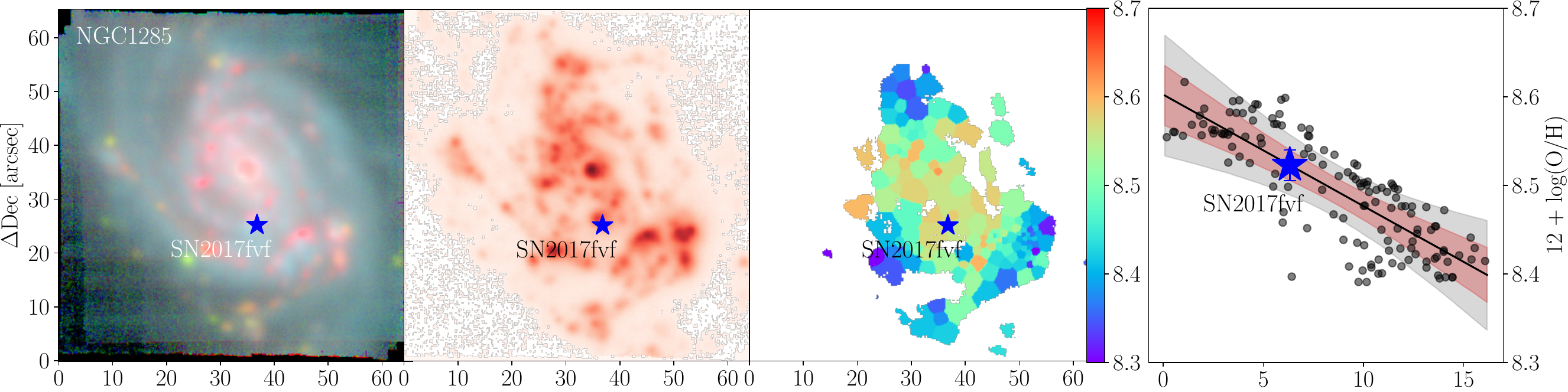}
    \includegraphics[width=1.0\linewidth]{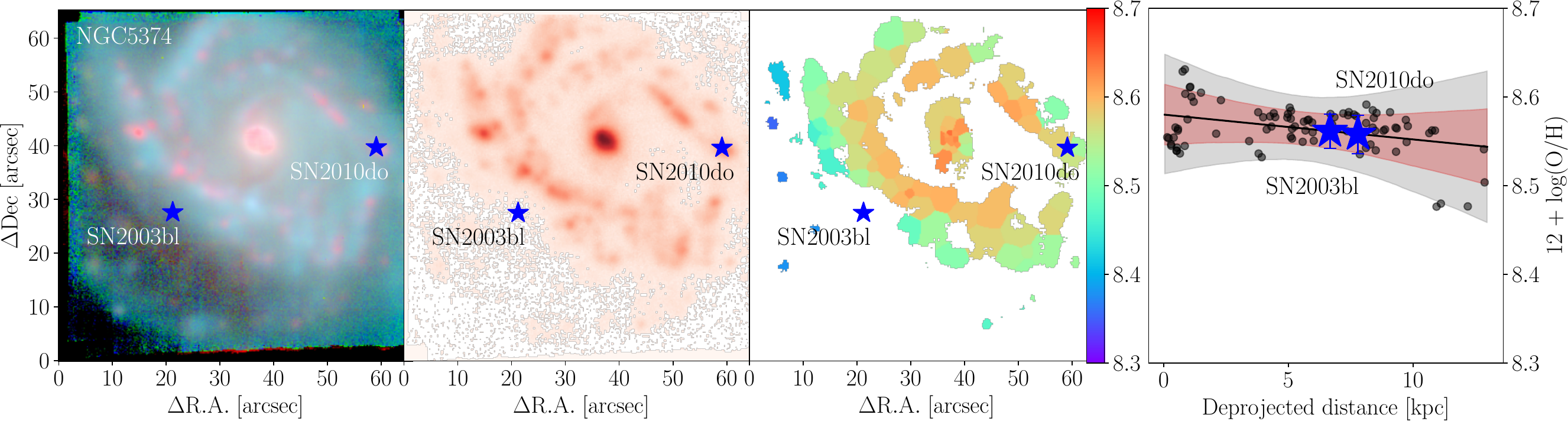}

    \caption{Example results of metallicity measurements for 7 SNe located in 4 host galaxies. Column 1: RGB images of host galaxies generated from MUSE datacube. The RGB components correspond to the cumulative fluxes from three spectral bands: 6550--6750 \AA, 4950--5150 \AA, and 4750--4950 \AA respectively. Column 2: H$\alpha$ flux maps generated by simulating narrowband filter (6548--6578 \AA) observations of the MUSE datacube. The continuum is fitted and subtracted using flux measurements from two adjacent wavelength bands: 6488--6518 \AA on the blue side and 6608--6638 \AA on the red side of the emission line.  The color scale is in arbitrary units. Column 3: Metallicity distribution maps derived with the strong-line method. Column 4: Metallicity gradient fitting results. Black dots represent metallicity measurements for individual bins. The solid line shows the Bayesian regression fit for the metallicity gradient, while the red and gray shaded regions indicate the 1$\sigma$ and 2$\sigma$ confidence intervals, respectively. The blue stars mark the SN locations.
}
    \label{fig:IC4397}
\end{figure*}
The typical measurement uncertainty is 0.18 dex for the strong-line method \citep{Marino2013A&A...559A.114M}. To reduce the metallicity uncertainties for the SNe, we used the galaxy metallicity gradient, calculated based on a large number of bins, to constrain the metallicity at the SN position. By using the spatial distribution characteristics of galaxy-wide metallicity, gradient fitting integrates information from multiple observation points, reduces the impact of local measurement uncertainties and enables safe extrapolation within a certain range. This effectively reduces the uncertainty in estimating metallicity at the SN position. Furthermore, some observations are made after the SN explosion, where the local spectra are contaminated by the SN’s light. The method of estimating the metallicity at the explosion site using metallicity gradients can effectively mitigate this contamination.

To calculate the metallicity gradient, we first removed the bins that do not correspond to star-forming regions using the Baldwin-Phillips-Terlevich
(BPT) diagram \citep{Baldwin1981}, adopting the maximum starburst line of \citet{Kewley2001}. Then for each Voronoi bin we calculated the deprojected distances to the galaxy center using the inclination and position angles from HyperLEDA; for some host galaxies, this information is not available and we derived the inclination and position angles by manually fitting the images. We fit the metallicity gradient using Bayesian regression, assuming Gaussian uncertainties for the individual metallicity measurements. The derived gradient was then used to estimate the metallicity at the SN position. In some circumstances, the SNe reside outside of, but not too far away from, the distance range, so we could safely extrapolate the gradient to derive the metallicity.

For some galaxies, it is difficult to fit a metallicity gradient, including the edge-on galaxies, for which we could not derive the deprojected distances, and those galaxies with too few Voronoi bins. In such cases, we calculated the metallicity from a local bin centered on the SN with a radius of 300 pc or the seeing-limited spatial resolution, whichever is larger. For SN2016hbb, SN2018eog, and SN2018dfh, we had to use a local bin to measure their metallicity, but the SNe were still very bright during the observations; therefore, we could not get an accurate metallicity measurement because of the SN contamination. These three were excluded from our analysis. 

\begin{figure}
    \centering
    \includegraphics[width=1\linewidth]{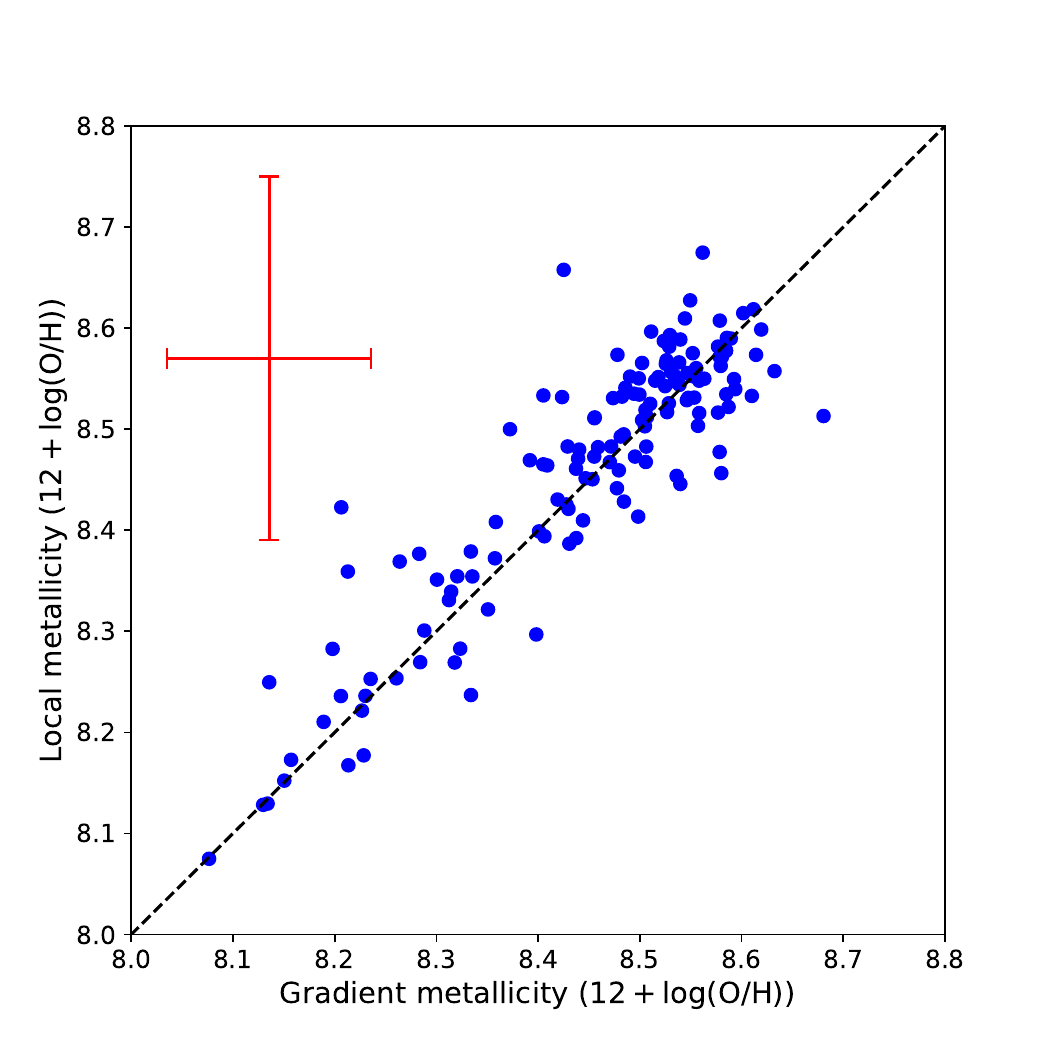}
    \caption{Comparison of SN metallicities obtained via the gradient method and directly from the local environment. The red error bars indicates the typical uncertainties: 0.18 for the local environment method and 0.10 (median) for the gradient method.}
    \label{fig:grad_local}
\end{figure}

Figure~\ref{fig:grad_local} compares the metallicity determined via the gradient method and those obtained directly through local environment. The two sets of measurements are consistent within uncertainties. However, the metallicity uncertainty derived via the gradient method (with a median value of 0.1 dex) is markedly smaller than that obtained directly from the local environment (0.18 dex). Moreover, the gradient approach effectively circumvents issues arising from explosion sites where the metallicity falls below the detection limit or is affected by contamination from SN light.

\section{Results and Discussion} \label{result}

\begin{figure}
    \centering
    \includegraphics[width=1.0\linewidth]{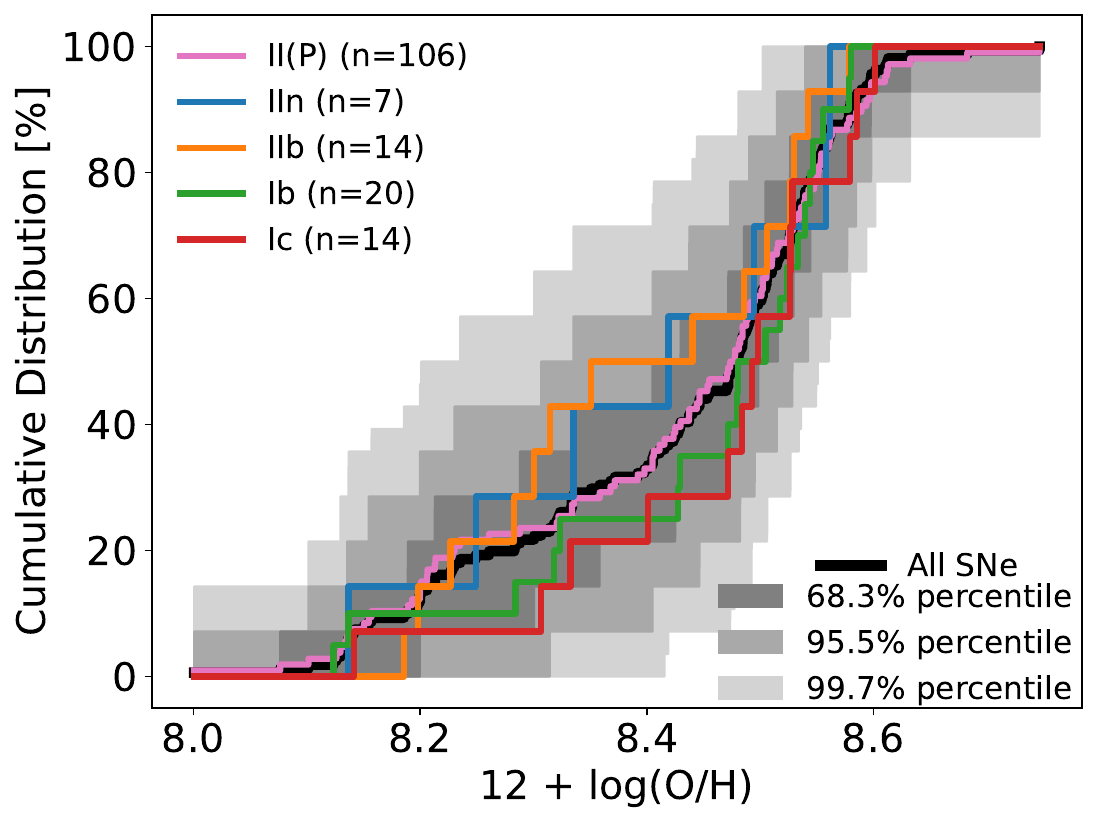}
    \caption{Cumulative metallicity distributions for different types of CCSNe. The black line represents the metallicity distribution for all SNe in the sample. The grey-shaded regions (from dark to light) indicate the 1$\sigma$, 2$\sigma$, and 3$\sigma$ uncertainties caused by the stochastic sampling effect for a $N$ = 14 subsample as estimated from our random resampling experiment.}
    \label{fig:cdf}
\end{figure}

With the method described above, we derived the metallicity for all SNe in our sample (listed in Table~\ref{table_snlist}). For example, Figure~\ref{fig:IC4397} displays the RGB composite images, H~\textsc{ii} regions, metallicity maps, and metallicity gradients of 4 host galaxies, with which we derived the metallicities for 7 SNe. Figure~\ref{fig:cdf} shows the cumulative metallicity distributions for all SNe and for different Types. For SN2014cw and SN2016dsb, the [N~\textsc{ii}] lines are below the detection limit, allowing only upper limits to be determined; therefore, they are not included in Figure~\ref{fig:cdf}. The metallicities span a range from 12 + log(O/H) = 8.1 to 8.7~dex. Assuming Gaussian measurement errors, we employed a multiple resampling approach to calculate the mean, median and standard deviation values of the metallicity distributions (the results are listed in Table~\ref{tab:metallicity_statistics}). The mean and median values are typically 8.4--8.5~dex, and the standard deviations are typically 0.17--0.20 dex. The differences among different SN types are very small. Type IIb and Type Ic have apparently the most different metallicity distributions, with mean (median) values of 8.39 (8.40) and 8.46 (8.51), respectively.

\begin{table}
    \centering
    \caption{Mean, median, and standard deviation values of 12+log(O/H) for different SN types. The errors originate from measurement uncertainties.} 
    \begin{tabular}{ccccc}
        \hline
        \hline
        SN Type &  Number & Mean& Median & Standard Deviation\\
        & & [dex] & [dex] & [dex] \\
        \hline
        II(P)   & 106&8.42 $\pm$ 0.01 & 8.47 $\pm$ 0.01 & 0.19   \\
        IIn  & 7&8.39 $\pm$ 0.05 & 8.42 $\pm$ 0.05 & 0.20\\
        IIb  & 14&8.39 $\pm$ 0.03 & 8.40 $\pm$ 0.04 & 0.18 \\
        Ib   & 20&8.44 $\pm$ 0.02 & 8.48 $\pm$ 0.02 & 0.17 \\
        Ic   & 14&8.46 $\pm$ 0.03 & 8.51 $\pm$ 0.03 & 0.17\\
        \hline
    \end{tabular}
    \label{tab:metallicity_statistics}
\end{table}

\subsection{Is there any significant metallicity difference among SN types?}



\begin{figure}
    \centering
    \includegraphics[width=1.0\linewidth]{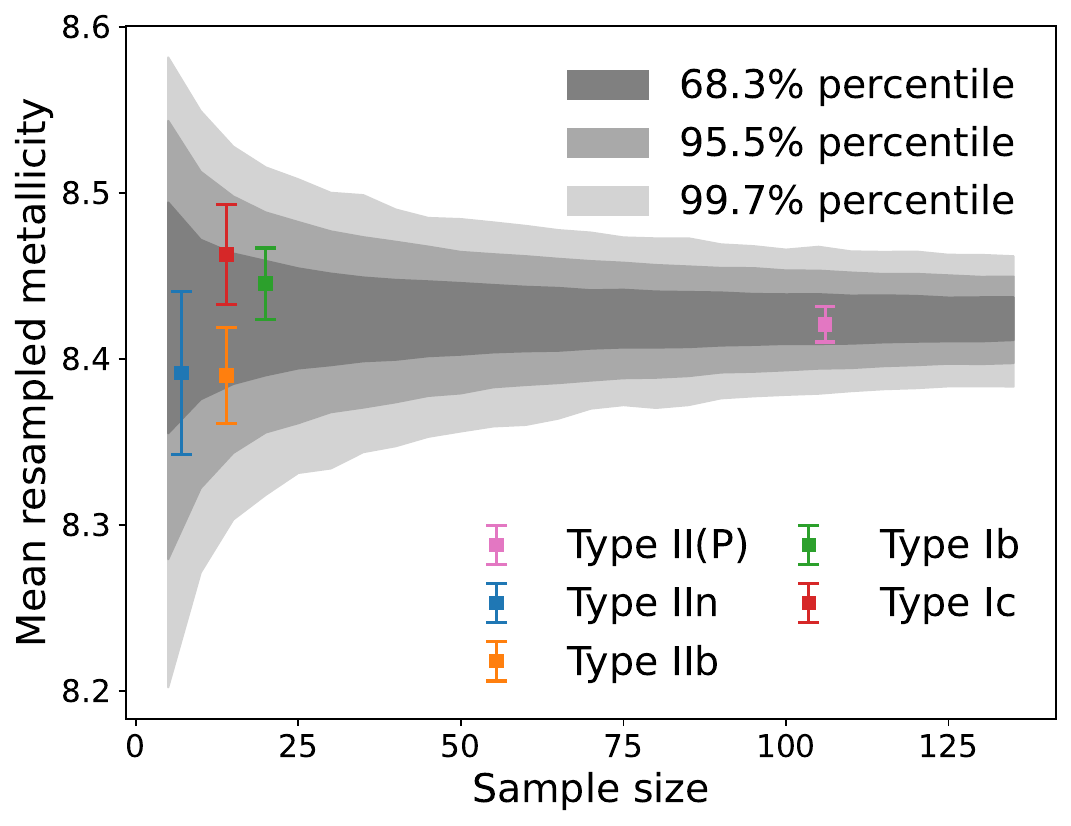}
    \caption{Data points: the number and mean metallicities for different types of CCSNe; the error bars are propagated from individual metallicity measurement uncertainties. Shaded regions: the 1$\sigma$, 2$\sigma$ and 3$\sigma$ (from dark to light) distributions of the mean values of randomly resampled SN metallicities from the full sample.}
    \label{fig:mean_resample}
\end{figure}

\begin{figure}
    \centering
    \includegraphics[width=1\linewidth]{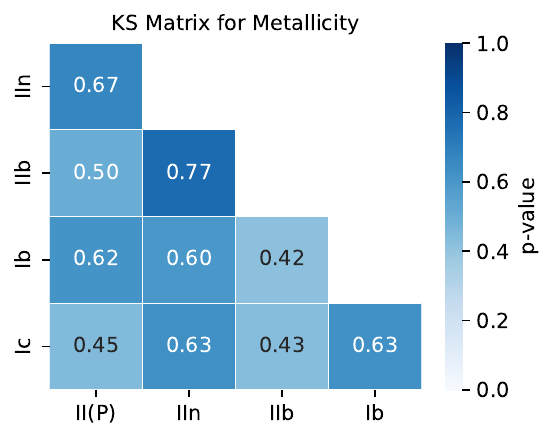}
    \includegraphics[width=1\linewidth]{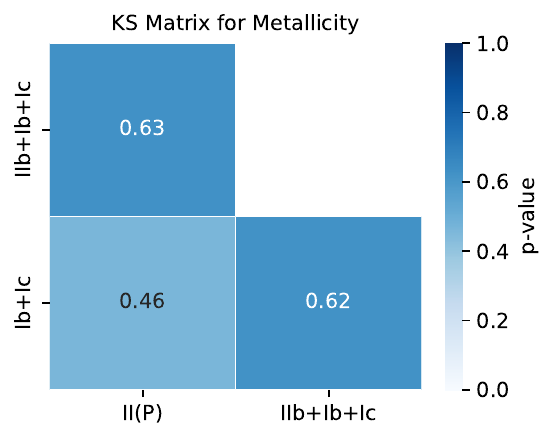}
    \caption{The $p$-value from KS test for each pair of SN types.}
    \label{fig:KS}
\end{figure}

For the derived metallicity distributions, we carried out an experiment to study whether the apparent difference among SN types is real or due to the stochastic sampling effect. We used the metallicity distribution of all SNe, regardless of types, as a reference distribution. We then randomly drew $N$ = 14 SNe (i.e. the number of SNe for Types IIb, and Ic in our sample) from the full sample and plotted their metallicity distribution. This process was repeated 10,000 times to show how the metallicity distributions vary due to the stochastic sampling effect. The results are shown in Figure~\ref{fig:cdf}. The stochastic sampling effect can cause a 1$\sigma$ uncertainty of $\sim$0.05~dex in the distributions. The metallicity distributions for different SN types are all consistent with the reference distribution within $\sim$1$\sigma$ uncertainties.

We also performed the above experiment by varying the number of randomly chosen SNe. Figure~\ref{fig:mean_resample} shows the probability distributions of the mean values of the resampled SN metallicities as a function of sample size.
For the Types II(P), IIn, IIb, Ib and Ic, the typical uncertainty for their mean metallicities caused by stochastic sampling is $\sim$0.05~dex, much larger than those propagated from metallicity measurement errors (Table~\ref{tab:metallicity_statistics}).
The measured mean metallicites for different SN types are all consistent with that of the reference distribution within 1$\sigma$ uncertainties. Therefore, the metallicity distributions of different SN types are not significantly different and are all consistent with being randomly drawn from the same reference distribution.

As an alternative method, we carried out a Kolmogorov-Smirnov (KS) test and calculated a $p$-value for each pair of SN types. In the KS test, the p-value assesses the degree of agreement  between two sample distributions. Typically, a $p$-value less than 0.05 indicates a statistically significant difference between the two samples; conversely, a $p$-value greater than 0.05 indicates insufficient evidence to reject the null hypothesis that the two samples are drawn from the same distribution.
The results are shown in Figure~\ref{fig:KS}. The $p$-values are generally very large, suggesting very weak metallicity differences among SN types. Even for the apparently most distinct Type IIb and Type Ic, the p-value is $\sim$0.4 and not small enough to indicate a significant metallicity difference between the two types. In addition, Type II(P) SNe do not exhibit significant differences compared to  the SESNe (IIb+Ib+Ic) grouped together.

\subsection{Comparison with previous results}

\cite{Sanders2012} studied the environments of a sample of SNe discovered by untargeted SN searches. They observed 75 Types IIb, Ib, Ic and Ic-BL SNe using the 6.5 m Magellan Telescopes at Las Campanas Observatory in Chile. They claimed a marginally significant difference between Type Ib and Type Ic SNe (with a $p$-value of $\sim$0.1 from KS test) and suggested that this difference may influence $\lesssim$30\% of stellar winds. This study relied on long-slit spectroscopy, however and was unable to spatially resolve the host galaxies.

\cite{Kuncarayakti2012Mass,Kuncarayakti2012su, Kuncarayakti2013, Kuncarayakti2013b,Kuncarayakti2015,Kuncarayakti2018} pioneered in using IFU spectroscopy to study SN environments. They investigated $\sim$100 SNe of different types based on observations with VLT (MUSE, VIMOS and SINFONI), Gemini-North (GMOS) and the Hawaii 2.2-m telescope (SNIFS). They found no significant metallicity differences among SN types \citep{Kuncarayakti2018}. By the time of their studies, however, most SNe were discovered by targeted searches and it is unclear whether this potential bias may influence their sample.

\citet{galbany2016,Galbany2018} compiled a large collection of SN host galaxies (i.e. the PISCO sample) based on IFU observations with the 3.5-m CAHA telescope at the Calar Alto Observatory in Spain. Their sample contained 272 SNe (including 120 Type~Ia SNe and 152 CCSNe) in 232 host galaxies. As noticed by themselves, most SNe in their sample were from targeted searches, therefore introducing a bias in the derived metallicity distributions. They also constructed a minimally biased sample from archival data and found that Type II(P) and Type Ic SNe display the highest metallicities while Type IIb and Type Ib SNe have lower metallicities. However, their KS test shows that this difference is not very significant.

\cite{pessi2023} conducted IFU observations with VLT/MUSE of a minimally biased sample of CCSNe discovered by the ASAS-SN survey (i.e. the AMUSING program). Their sample included a total of 112 CCSNe and they did not find any significant metallicity differences among the SN types. However, most SNe in their sample are of Type II and very few are of the other types (9 IIn, 7 IIb, 7 Ib, 4 Ic, 3 Ibn, 2 Ic-BL). So their result may suffer more from the stochastic sampling effect.

In summary, the previous studies have not found any significant metallicity differences among the main CCSN types (II(P), IIn, IIb, Ib and Ic). Now based on a larger and minimally biased sample with IFU observations, our study further confirms this conclusion. The typical uncertainty caused by stochastic sampling is narrowed down to $\sim$0.05~dex and our careful analysis shows that all the SN types are consistent within a $\sim$1$\sigma$ level.

\subsection{The role of metallicity in SN progenitors}

In the single-star progenitor channel, SESNe originate from massive WR stars, whose outer envelopes are stripped by their stellar winds \citep{conti1978}. The strength of line-driven wind is very sensitive to metallicity \citep{Castor1975,Kudritzki2000,Vink2001} and one may expect an increasing trend in metallicity for Types IIP, IIb, Ib and Ic with increasing degrees of envelope stripping. However, our result shows no significant metallicity difference between these SN types. It is possible that the binary progenitor channel dominates the origin of most SESNe. In this case, the dependence on metallicity is minimal, while binary parameters (such as orbital separation and secondary-to-primary mass ratio) exert a greater influence. This conclusion is consistent with those based on SN fraction \citep{Smith2011}, direct progenitor/companion detections \citep{Crockett2008,Folatelli_2015,Niu202417gkk,sun20202014c}, nebular spectroscopy \citep{Maeda2006,Maeda2014,Maeda2015,Fang_2018}, and light curve modeling \citep{Lyman2016,Taddia2018,Woosley_2021}.

Although not as sensitive as stellar wind, the stripping of envelopes via binary interaction is not independent of metallicity.
Recent studies suggest that Roche-lobe stripping may become significantly less efficient at low metallicities \citep{gotberg2017}. In high-metallicity stars, the greater opacity in the outer layers can trap radiation and the higher radiative pressure can help to expel the hydrogen envelope. In low-metallicity stars, however, the lower opacity allows radiation to escape more easily, thus reducing the radiative pressure and keeping the hydrogen envelope cooler and more tightly bound; therefore, it is easier for a low-metallicity mass donor to retain a significant hydrogen envelope after binary interaction, and this will result in a Type IIb, instead of a Type Ib, SN explosion. In our result, Type~IIb SNe seem to have the lowest metallicities, but this difference is not significant enough to support this hypothesis. Future studies with even larger samples will be necessary to reveal the possible metallicity differences among SN types.

\section{Summary and conclusions} \label{summary}

In this work, we studied the metallicity of CCSNe based on a large and minimally biased sample with IFU observations. We carefully selected nearby CCSNe with archival VLT/MUSE data by considering the potential biases introduced by SN discovery and data availability. The final sample contains 166 CCSNe at a redshift of $z \leq$ 0.02 discovered by untargeted SN searches, covering the main CCSN types of II(P), IIn, IIb, Ib and Ic. Such a sample is representative of the SN population in the local Universe and is to date the largest sample for SN metallicity studies based on IFU observations.

For each SN host galaxy, we derived the spatially-resolved metallicity map with the strong-line method and estimated the SN metallicity with the galaxy metallicity gradient. The derived metallicities range from 12 + log(O/H) = 8.1 to 8.7 dex; for different SN types, the mean and median values are typically 8.4--8.5~dex, and the standard deviations are typically 0.17--0.20 dex.

With a random resampling experiment and a KS test, we show that there is no significant metallicity difference among the analyzed SN types. They can all be considered as being drawn randomly from the same reference distribution. With our large sample, the uncertainty caused by the stochastic sampling effect is narrowed down to $\sim$0.05~dex, and the metallicity distributions of different SN types are all consistent within $\sim$1$\sigma$ uncertainties.

In the single-star progenitor channel, where mass loss is dominated by metallicity-dependent line-driven wind, we expect an increasing trend of metalliciy for IIP $\rightarrow$ IIb $\rightarrow$ Ib $\rightarrow$ Ic with increasing degrees of envelope stripping. However, our result suggests that metallicity plays a very minor role in the origin of SESNe. It is possible that most SESNe are from the binary progenitor channel, where the final fate of massive stars is insensitive to metallicity but is primarily determined by the binary parameters (e.g. secondary-to-primary mass ratio, binary separation).

Some theoretical studies suggest that Robe-lobe stripping becomes less efficient at low metallicities such that the progenitor may retain a significant hydrogen envelope and result in a Type~IIb SN explosion. In our results, although the metallicities of Type~IIb SNe are lower by more than 1$\sigma$ uncertainties, they are still consistent with the reference distribution within 2$\sigma$ uncertainties. Future studies with even larger samples will be necessary to reveal the possible metallicity differences among SN types.

\section*{acknowledgments}
This work is supported by the Strategic Priority Research Program of the Chinese Academy of Sciences, Grant No. XDB0550300. NCS’s research is funded by the NSFC grants No. 12303051 and No. 12261141690 and ZXN acknowledges support from the NSFC through grant No. 12303039. JFL acknowledges support from the NSFC through grants No. 11988101 and No. 11933004 and from the New Cornerstone Science Foundation through the New Cornerstone Investigator Program and the XPLORER PRIZE.

\section*{Data Availability}
Based on observations collected at the European Southern Observatory under ESO programme(s): 60.A-9319(A), 097.B-0640(A), 106.210Z.008, 096.B-0309(A), 106.210Z.009, 098.B-0193(A), 110.23ZH.001, 097.B-0518(A), 111.24VQ.001, 60.A-9301(A), 095.B-0686(A), 1100.B-0651(B), 0101.D-0748(A), 095.D-0172(A), 0104.D-0503(A), 104.20VC.001, 0103.D-0440(A), 0103.A-0637(A), 096.D-0263(A), 106.2155.001, 0101.B-0368(B), 1100.B-0651(A), 096.D-0296(A), 1100.B-0651(D), 0104.B-0404(A), 097.B-0165(A), 108.229J.001, 095.B-0042(A), 097.B-0041(A), 0100.D-0341(A), 094.B-0733(B), 108.21ZY.008, 0101.A-0282(A), 099.A-0870(A), 106.2104.001, 0102.B-0794(A), 111.24UM.001, 0101.B-0706(A), 110.24AS.002, 0100.B-0116(A), 097.D-0408(A), 099.B-0242(A), 099.D-0022(A), 108.229G.001, 094.B-0298(A)



\bibliographystyle{mnras}
\bibliography{example} 




\appendix

\section{SN Data with Metallicity}

\textbf{}
\onecolumn
\begin{center}
    \captionsetup{justification=raggedright, singlelinecheck=false} 
    \begin{longtable}{lccccccccc}
    \caption{SN Data with Metallicity. PA: Position angle of host galaxy; $i$: Inclination angle of host galaxy. 12+log(O/H): Oxygen abundance at SN location.} \label{table_snlist} \\
    \hline
    \hline
    \makecell{Name} & \makecell{Type} & \makecell{Host galaxy} & \makecell{Redshift} & \makecell{PA \\ {[deg]}} & \makecell{$i$ \\ {[deg]}} & \makecell{12+log(O/H) \\ {grad. [dex]}} & \makecell{12+log(O/H) \\ {local [dex]}}& \makecell{Calibration} \\
    \midrule
    \endfirsthead
    \caption[]{SN Data with Metallicity (continued)} \\
    \toprule

    \endhead

    \multicolumn{9}{r}{Continued on next page} \\
    \bottomrule
    \endfoot
    \bottomrule
    \endlastfoot

ASASSN-14dl & II & ESO 506-G4 & 0.0134 & 88.4 & 67.2 & 8.55(+0.05/$-$0.05) & 8.53 & O3N2 \\
ASASSN-14dp & II & ESO 319-G15 & 0.0092 & 81.5 & 54.2 & 8.15(+0.09/$-$0.09) & -- & O3N2 \\
ASASSN-14dq & II & UGC 11860 & 0.0104 & 133.0 & 74.7 & 8.21(+0.08/$-$0.08) & 8.42 & O3N2 \\
ASASSN-14ha & II & NGC 1566 & 0.0050 & 44.2 & 49.1 & 8.58(+0.02/$-$0.02) & 8.52 & O3N2 \\
ASASSN-14ma & II & SDSS J235509.00+101252.9 & 0.0137 & 89.1 & 29.2 & 8.29(+0.04/$-$0.04) & 8.30 & O3N2 \\
ASASSN-15bb & II & ESO 381-IG48 & 0.0159 & 110.6 & 59.1 & 8.14(+0.06/$-$0.06) & 8.25 & O3N2 \\
ASASSN-15fi & II & MRK 884 & 0.0172 & 45.5 & 40.0 & 8.13(+0.01/$-$0.01) & 8.13 & O3N2 \\
ASASSN-15fz & II & NGC 5227 & 0.0175 & 161.1 & 32.8 & 8.52(+0.06/$-$0.06) & 8.54 & O3N2 \\
ASASSN-15jp & II & NGC 3157 & 0.0095 & 39.1 & 80.4 & 8.47(+0.05/$-$0.05) & -- & O3N2 \\
ASASSN-15ln & II & UGC 546 & 0.0150 & 3.3 & 77.8 & 8.21(+0.07/$-$0.07) & 8.36 & O3N2 \\
ASASSN-15lx & II & ESO 47-G4 & 0.0126 & 90.5 & 48.7 & 8.21(+0.05/$-$0.05) & 8.24 & O3N2 \\
ASASSN-15oz & II & HIPASS J1919-33 & 0.0069 & -- & -- & -- & 8.66 & O3N2 \\
ASASSN-15qh & II & ESO 534-G024 & 0.0102 & 112.0 & 55.6 & 8.41(+0.09/$-$0.09) & -- & O3N2 \\
ASASSN-16ab & II & CGCG 012-116 & 0.0043 & 49.0 & 52.5 & 8.24(+0.05/$-$0.05) & 8.25 & O3N2 \\
ASASSN-19kz & II & NGC 2207 & 0.0091 & 115.6 & 58.2 & 8.52(+0.03/$-$0.03) & -- & O3N2 \\
AT2018bbl & II & NGC 7421 & 0.0060 & 80.6 & 36.2 & 8.59(+0.07/$-$0.07) & 8.54 & O3N2 \\
PS15aaa & II & IC 564 & 0.0190 & 68.2 & 77.3 & 8.51(+0.06/$-$0.06) & 8.52 & O3N2 \\
PS15afa & II & NGC 3404 & 0.0150 & 81.3 & 86.7 & 8.60(+0.14/$-$0.14) & -- & O3N2 \\
PTF09gpn & II & Anonymous & 0.0150 & -- & -- & -- & 8.32 & O3N2 \\
SMT16atf & II & PGC098793 & 0.0140 & 110.0 & 0.0 & 8.41(+0.06/$-$0.06) & 8.39 & O3N2 \\
SN1998dl & II & NGC 1084 & 0.0044 & 39.9 & 49.9 & 8.46(+0.01/$-$0.01) & 8.51 & O3N2 \\
SN1999dh & II & IC 211 & 0.0110 & 56.0 & 64.7 & 8.40(+0.02/$-$0.02) & 8.30 & O3N2 \\
SN2001J & II & UGC 4729 & 0.0130 & 85.0 & 35.2 & 8.36(+0.05/$-$0.05) & 8.37 & O3N2 \\
SN2003E & II & ESO 485-G004 & 0.0149 & 142.9 & 90.0 & -- & 8.31 & O3N2 \\
SN2003ao & II & NGC 2993 & 0.0081 & 93.7 & 35.8 & 8.44(+0.01/$-$0.01) & 8.47 & O3N2 \\
SN2004F & II & NGC 1285 & 0.0175 & 8.1 & 59.3 & 8.51(+0.02/$-$0.02) & 8.60 & O3N2 \\
SN2004ci & II & NGC 5980 & 0.0140 & 14.5 & 76.4 & 8.59(+0.03/$-$0.03) & 8.52 & O3N2 \\
SN2005H & II & NGC 838 & 0.0128 & 77.2 & 49.8 & 8.53(+0.01/$-$0.01) & 8.55 & O3N2 \\
SN2005Z & II & NGC 3363 & 0.0190 & 179.2 & 45.3 & 8.62(+0.06/$-$0.06) & 8.60 & O3N2 \\
SN2006be & II & IC 4582 & 0.0071 & 172.1 & 83.1 & 8.37(+0.14/$-$0.14) & 8.50 & O3N2 \\
SN2006ca & II & UGC 11214 & 0.0088 & 175.0 & 16.5 & 8.41(+0.06/$-$0.06) & 8.46 & O3N2 \\
SN2006cx & II & NGC 7316 & 0.0185 & 66.0 & 32.9 & 8.52(+0.02/$-$0.02) & 8.55 & O3N2 \\
SN2007rw & II & UGC 7798 & 0.0086 & 57.2 & 56.0 & 8.31(+0.06/$-$0.06) & 8.33 & O3N2 \\
SN2008V & II & NGC 1591 & 0.0137 & 29.4 & 56.8 & 8.55(+0.04/$-$0.04) & 8.53 & O3N2 \\
SN2008aw & II & NGC 4939 & 0.0104 & 7.4 & 70.1 & 8.68(+0.19/$-$0.19) & 8.51 & O3N2 \\
SN2008fq & II & NGC 6907 & 0.0106 & 57.7 & 37.5 & 8.58(+0.01/$-$0.01) & 8.53 & O3N2 \\
SN2009H & II & NGC 1084 & 0.0047 & 39.9 & 49.9 & 8.46(+0.01/$-$0.01) & 8.51 & O3N2 \\
SN2009K & II & NGC 1620 & 0.0117 & 22.9 & 81.2 & 8.61(+0.11/$-$0.11) & 8.57 & N2 \\
SN2009au & II & ESO 443-G21 & 0.0094 & 159.6 & 79.0 & 8.46(+0.02/$-$0.02) & 8.48 & O3N2 \\
SN2009dq & II & IC 2554 & 0.0046 & 4.1 & 70.8 & 8.58(+0.01/$-$0.01) & 8.58 & O3N2 \\
SN2010F & II & NGC 3120 & 0.0093 & 6.2 & 47.5 & 8.51(+0.07/$-$0.07) & 8.51 & O3N2 \\
SN2010K & II & A120246+0224 & 0.0200 & -- & -- & -- & 8.13 & O3N2 \\
SN2010cl & II & MCG -02-25-20 & 0.0091 & 126.2 & 85.5 & 8.56(+0.11/$-$0.11) & -- & O3N2 \\
SN2012cc & II & NGC 4419 & -0.0009 & 132.7 & 84.7 & 8.59(+0.06/$-$0.06) & -- & O3N2 \\
SN2012ga & II & NGC 6976 & 0.0200 & 164.9 & 27.1 & 8.50(+0.06/$-$0.06) & 8.50 & O3N2 \\
SN2013ej & II & NGC 628 & 0.0022 & 25.0 & 19.8 & 8.51(+0.04/$-$0.04) & -- & O3N2 \\
SN2014V & II & NGC 3905 & 0.0193 & 62.5 & 48.7 & 8.53(+0.04/$-$0.04) & 8.56 & O3N2 \\
SN2014cw & II & PGC 68414 & 0.0060 & -- & -- &-- &< 8.28&N2 \\
SN2014ay & II & UGC 11037 & 0.0104 & 52.2 & 90.0 & -- & 8.49 & O3N2 \\
SN2014cy & II & NGC 7742 & 0.0055 & 165.0 & 16.8 & 8.55(+0.01/$-$0.01) & 8.53 & O3N2 \\
SN2014dw & II & NGC 3568 & 0.0082 & 7.0 & 67.0 & 8.48(+0.02/$-$0.02) & 8.49 & O3N2 \\
SN2015ay & II & UGC 722 & 0.0140 & 136.9 & 90.0 & -- & 8.20 & O3N2 \\
SN2016adl & II & GALEXASC J115155.68-132459.3 & 0.0070 & -- & -- & -- & 8.07 & O3N2 \\
SN2016aqf & II & NGC 2101 & 0.0040 & 94.0 & 69.1 & 8.23(+0.06/$-$0.06) & 8.18 & O3N2 \\
SN2016ase & II & ESO 504- G 009 & 0.0150 & 123.1 & 47.0 & 8.14(+0.15/$-$0.16) & -- & O3N2 \\
SN2016bev & II & ESO 560-G013 & 0.0110 & 138.8 & 90.0 & -- & 8.37 & O3N2 \\
SN2016blz & II & SDSS J154029.29+005437.4 & 0.0110 & 0.8 & 44.6 & 8.19(+0.06/$-$0.06) & 8.21 & O3N2 \\
SN2016bsb & II & Anonymous & 0.0200 & -- & -- & -- & 8.17 & O3N2 \\
SN2016cyk & II & 2MASX J13024397-2656276 & 0.0161 & 70.0 & 55.8 & 8.56(+0.04/$-$0.04) & 8.55 & O3N2 \\
SN2016hgm & II & NGC 493 & 0.0080 & 59.9 & 74.6 & 8.44(+0.08/$-$0.08) & 8.39 & O3N2 \\
SN2016hmq & II & PGC146262 & 0.0174 & 28.5 & 73.5 & 8.48(+0.07/$-$0.07) & 8.44 & O3N2 \\
SN2016iyz & II & IC 2151 & 0.0104 & 93.4 & 61.5 & 8.49(+0.03/$-$0.03) & 8.55 & O3N2 \\
SN2016zb & II & MCG -03-25-015 & 0.0140 & 120.2 & 18.6 & 8.19(+0.25/$-$0.25) & -- & O3N2 \\
SN2017ahn & II & NGC3318 & 0.0090 & 79.4 & 59.8 & 8.50(+0.02/$-$0.02) & 8.47 & O3N2 \\
SN2017ahn & II & NGC3318 & 0.0090 & 79.4 & 59.8 & 8.46(+0.06/$-$0.06) & 8.47 & O3N2 \\
SN2017auf & II & MCG -02-13-038 & 0.0133 & 111.3 & 73.6 & 8.61(+0.06/$-$0.06) & 8.53 & O3N2 \\
SN2017faa & II & IC 4224 & 0.0180 & 99.3 & 84.2 & 8.39(+0.08/$-$0.08) & 8.47 & O3N2 \\
SN2017fbq & II & 2MASX J19334551-6058022 & 0.0150 & 161.0 & 81.1 & 8.33(+0.06/$-$0.06) & 8.24 & O3N2 \\
SN2017fbu & II & IC 211 & 0.0109 & 56.0 & 64.7 & 8.40(+0.02/$-$0.02) & -- & O3N2 \\
SN2017ffq & II & 2MASX J17401447-5825586 & 0.0127 & 140.8 & 74.4 & 8.45(+0.05/$-$0.05) & 8.45 & O3N2 \\
SN2017fqk & II & NGC 1137 & 0.0150 & 16.1 & 59.5 & 8.48(+0.11/$-$0.11) & -- & O3N2 \\
SN2017fqo & II & NGC 716 & 0.0150 & 59.0 & 75.9 & 8.43(+0.19/$-$0.19) & 8.42 & O3N2 \\
SN2017ggw & II & ESO-246-G-21 & 0.0180 & 140.7 & 52.4 & 8.51(+0.05/$-$0.05) & -- & O3N2 \\
SN2017gmr & II & NGC0988 & 0.0050 & 119.6 & 69.1 & 8.49(+0.03/$-$0.03) & -- & O3N2 \\
SN2017grn & II & IC1498 & 0.0180 & 2.9 & 90.0 & -- & 8.59 & O3N2 \\
SN2017hxv & II & ESO 466- G 004 & 0.0160 & 134.4 & 41.3 & 8.60(+0.07/$-$0.07) & -- & O3N2 \\
SN2017jmk & II & NGC7541 & 0.0095 & 101.6 & 74.8 & 8.48(+0.02/$-$0.02) & 8.53 & O3N2 \\
SN2017pn & II & PGC959170 & 0.0140 & 38.0 & 62.4 & 8.20(+0.07/$-$0.07) & -- & O3N2 \\
SN2018ant & II & MCG -02-22-22 & 0.0197 & 70.0 & 90.0 & -- & 8.68 & O3N2 \\
SN2018bl & II & ESO 18-G9 & 0.0180 & 50.0 & 34.3 & 8.54(+0.06/$-$0.06) & 8.45 & O3N2 \\
SN2018cuf & II & IC5092 & 0.0108 & 26.9 & 28.6 & 8.61(+0.19/$-$0.19) & -- & O3N2 \\
SN2018cvn & II & ESO 476- G 016 & 0.0190 & 141.1 & 59.6 & 8.45(+0.19/$-$0.20) & 8.45 & O3N2 \\
SN2018dfg & II & NGC5468 & 0.0095 & 109.2 & 21.1 & 8.56(+0.04/$-$0.04) & -- & O3N2 \\
SN2018evy & II & NGC 6627 & 0.0180 & 74.5 & 26.9 & 8.53(+0.03/$-$0.03) & 8.55 & O3N2 \\
SN2018fit & II & CGCG 431-062 & 0.0140 & 130.6 & 81.5 & 8.54(+0.27/$-$0.27) & -- & O3N2 \\
SN2018hyw & II & UGC 4344  & 0.0168 & 89.4 & 27.7 & 8.42(+0.07/$-$0.08) & 8.53 & O3N2 \\
SN2018ivc & II & NGC1068 & 0.0038 & 72.7 & 34.7 & 8.48(+0.00/$-$0.00) & 8.43 & O3N2 \\
SN2018kcw & II & IC 5179 & 0.0120 & 60.6 & 62.2 & 8.55(+0.01/$-$0.01) & 8.58 & O3N2 \\
SN2018lab & II & IC2163 & 0.0092 & 102.6 & 78.2 & 8.53(+0.01/$-$0.01) & 8.56 & O3N2 \\
SN2018pq & II & IC 3896A & 0.0060 & 105.0 & 48.4 & 8.53(+0.16/$-$0.15) & -- & O3N2 \\
SN2019dxd & II & NGC 3464 & 0.0125 & 110.8 & 50.8 & 8.54(+0.05/$-$0.05) & 8.59 & O3N2 \\
SN2019hyk & II & IC 4397 & 0.0147 & 160.1 & 48.3 & 8.50(+0.07/$-$0.07) & -- & O3N2 \\
SN2019ltw & II & CGCG 137-076 & 0.0160 & 59.0 & 25.4 & 8.44(+0.05/$-$0.05) & 8.46 & O3N2 \\
SN2019tua & II & UGC 11860 & 0.0104 & 133.0 & 74.7 & 8.16(+0.07/$-$0.06) & 8.17 & O3N2 \\
SN2019xis & II & Anonymous & 0.0050 & -- & -- & -- & 8.15 & O3N2 \\
SN2020aqe & II & NGC 3836 & 0.0123 & 137.7 & 39.8 & 8.41(+0.02/$-$0.02) & 8.47 & O3N2 \\
SN2020aze & II & NGC3318 & 0.0090 & 79.4 & 59.8 & 8.55(+0.05/$-$0.05) & 8.55 & O3N2 \\
SN2020jfo & II & M61 & 0.0050 & 162.0 & 18.1 & 8.58(+0.05/$-$0.05) & 8.57 & O3N2 \\
SN2020llx & II & NGC 7140 & 0.0099 & 17.4 & 49.6 & 8.55(+0.08/$-$0.08) & 8.55 & O3N2 \\
SN2021abkm & II & NGC 6627 & 0.0176 & 74.5 & 26.9 & 8.58(+0.11/$-$0.11) & 8.58 & O3N2 \\
SN2021agdm & II & ESO 61-8 & 0.0114 & 106.8 & 78.0 & 8.45(+0.12/$-$0.12) & -- & O3N2 \\
SN2021zgm & II & UGC 11289 & 0.0133 & 1.0 & 53.7 & 8.61(+0.18/$-$0.18) & -- & O3N2 \\
SN2022aau & II & NGC1672 & 0.0044 & 154.9 & 28.9 & 8.55(+0.00/$-$0.00) & 8.63 & O3N2 \\
SN2022acko & II & NGC1300 & 0.0053 & 104.6 & 61.8 & 8.63(+0.10/$-$0.11) & 8.56 & O3N2 \\
SN2022mmr & II & IC 1498 & 0.0173 & 2.9 & 90.0 & -- & 8.59 & N2 \\
SN2022wsp & II & NGC 7448 & 0.0073 & 170.5 & 70.1 & 8.41(+0.01/$-$0.01) & 8.53 & O3N2 \\
SN2023dpj & II & NGC 5135 & 0.0137 & 126.4 & 24.8 & 8.51(+0.02/$-$0.02) & 8.47 & O3N2 \\
SN2023ijd & II & NGC 4568 & 0.0074 & 28.6 & 67.5 & 8.51(+0.02/$-$0.02) & -- & O3N2 \\
SN2023rve & II & NGC 1097 & 0.0040 & 133.9 & 54.8 & 8.75(+0.26/$-$0.25) & -- & O3N2 \\
SN2024jlf & II & NGC5690 & 0.0058 & 145.1 & 75.9 & 8.47(+0.06/$-$0.06) & 8.53 & O3N2 \\
ASASSN-14iz & IIP & ESO 462-G9 & 0.0193 & 162.3 & 58.8 & 8.48(+0.15/$-$0.15) & 8.57 & N2 \\
ASASSN-15kz & IIP & IC 4303 & 0.0080 & 70.7 & 59.1 & 8.23(+0.04/$-$0.04) & 8.24 & O3N2 \\
ASASSN-15ng & IIP & ESO 221-G12 & 0.0098 & 164.3 & 90.0 & -- & 8.34 & O3N2 \\
ASASSN-16at & IIP & UGC 8041 & 0.0044 & 168.3 & 54.0 & 8.32(+0.07/$-$0.07) & 8.35 & O3N2 \\
SN1999br & IIP & NGC 4900 & 0.0032 & 135.0 & 19.0 & 8.43(+0.02/$-$0.02) & 8.39 & O3N2 \\
SN2003bl & IIP & NGC 5374 & 0.0146 & 45.0 & 36.9 & 8.54(+0.02/$-$0.02) & 8.61 & N2 \\
SN2003bn & IIP & 2MASX J10023529-2110531 & 0.0128 & 98.0 & 74.6 & 8.36(+0.06/$-$0.06) & 8.41 & O3N2 \\
SN2003hg & IIP & NGC 7771 & 0.0143 & 68.0 & 66.7 & 8.58(+0.02/$-$0.02) & 8.58 & O3N2 \\
SN2012bu & IIP & NGC 3449 & 0.0109 & 145.8 & 90.0 & -- & 8.49 & O3N2 \\
SN2015W & IIP & UGC 3617 & 0.0130 & 8.5 & 49.0 & 8.10(+0.18/$-$0.18) & -- & O3N2 \\
SN2016B & IIP & CGCG 012-116 & 0.0043 & 49.0 & 52.5 & 8.26(+0.06/$-$0.06) & 8.25 & O3N2 \\
SN2016I & IIP & UGC 09450 & 0.0149 & 49.0 & 90.0 & -- & 8.12 & N2 \\
SN2016L & IIP & UGCA 397 & 0.0090 & 120.0 & 19.0 & 8.20(+0.05/$-$0.05) & -- & O3N2 \\
SN2016blb & IIP & 2MASX J11372059-0454450 & 0.0180 & 168.0 & 67.5 & 8.33(+0.06/$-$0.06) & 8.38 & O3N2 \\
SN2016cok & IIP & M66 & 0.0020 & 168.2 & 67.5 & 8.59(+0.03/$-$0.03) & 8.59 & O3N2 \\
SN2016hvu & IIP & NGC 7316 & 0.0185 & 66.0 & 32.9 & 8.44(+0.02/$-$0.02) & 8.41 & O3N2 \\
SN2017aym & IIP & NGC 5690 & 0.0058 & 145.1 & 75.9 & 8.51(+0.05/$-$0.05) & 8.53 & O3N2 \\
SN2017ejx & IIP & NGC 2993 & 0.0081 & 93.7 & 35.8 & 8.47(+0.01/$-$0.01) & 8.47 & O3N2 \\
SN2017fem & IIP & IC 4452 & 0.0140 & 77.8 & 20.6 & 8.50(+0.03/$-$0.03) & 8.51 & O3N2 \\
SN2017fvf & IIP & NGC 1285 & 0.0170 & 8.1 & 59.3 & 8.50(+0.02/$-$0.02) & 8.57 & O3N2 \\
SN2017fvr & IIP & UGC 3165 & 0.0130 & 135.0 & 61.0 & 8.42(+0.07/$-$0.07) & -- & O3N2 \\
SN2017gry & IIP & ESO 155-G36 & 0.0193 & 171.9 & 82.4 & 8.54(+0.05/$-$0.05) & 8.54 & O3N2 \\
SN2017ivu & IIP & NGC 5962 & 0.0065 & 106.3 & 51.4 & 8.00(+0.18/$-$0.18) & -- & O3N2 \\
SN2018cho & IIP & IC 4 & 0.0167 & 12.0 & 45.6 & 8.56(+0.04/$-$0.04) & 8.55 & O3N2 \\
SN2018yo & IIP & UGC 7840 & 0.0130 & 73.4 & 57.6 & 8.37(+0.09/$-$0.09) & -- & O3N2 \\
ASASSN-14fd & IIn & PGC 43070 & 0.0154 & 16.0 & 51.5 & 8.34(+0.07/$-$0.07) & 8.35 & O3N2 \\
ASASSN-15hs & IIn & 2MASX J15333488-7807258 & 0.0091 & 177.3 & 39.9 & 8.56(+0.03/$-$0.03) & 8.52 & O3N2 \\
ASASSN-16jt & IIn & ESO 344-G021 & 0.0108 & 58.0 & 67.3 & 8.56(+-0.06/$-$0.16) & 8.67 & O3N2 \\
SN1997bs & IIn & NGC 3627 & 0.0019 & 168.2 & 67.5 & 8.59(+0.04/$-$0.04) & 8.55 & O3N2 \\
SN2013fc & IIn & ESO 154-G10 & 0.0186 & 87.9 & 35.5 & 8.65(+0.05/$-$0.05) & -- & O3N2 \\
SN2015bf & IIn & NGC 7653 & 0.0142 & 172.5 & 31.0 & 8.49(+0.18/$-$0.18) & 8.54 & O3N2 \\
SN2016aiy & IIn & ESO 323-G084 & 0.0100 & 7.0 & 77.7 & 8.25(+0.18/$-$0.18) & -- & O3N2 \\
SN2016eso & IIn & ESO 422- G 019 & 0.0170 & 148.9 & 62.5 & 8.14(+0.32/$-$0.08) & -- & O3N2 \\
SN2021aefs & IIn & NGC 3836 & 0.0123 & 137.7 & 39.8 & 8.42(+0.01/$-$0.01) & 8.43 & O3N2 \\
ASASSN-14az & IIb & PGC 1101367 & 0.0067 & 12.0 & 68.8 & 8.20(+0.13/$-$0.13) & 8.28 & O3N2 \\
ASASSN-15bd & IIb & SDSS J155438.39+163637.6 & 0.0079 & 89.1 & 90.0 & -- & 8.19 & O3N2 \\
ASASSN-15tu & IIb & 2MASX J22340166-3223490 & 0.0126 & 65.0 & 38.6 & 8.35(+0.06/$-$0.06) & 8.32 & O3N2 \\
PS15apj & IIb & NGC 6641 & 0.0140 & 100.0 & 29.9 & 8.51(+0.06/$-$0.06) & 8.48 & O3N2 \\
SN2008aq & IIb & MCG -02-33-20 & 0.0080 & 175.0 & 90.0 & -- & 8.14 & O3N2 \\
SN2014cl & IIb & IC 217 & 0.0063 & 35.1 & 82.6 & 8.27(+0.28/$-$0.28) & -- & O3N2 \\
SN2015bi & IIb & VV 839 & 0.0160 & 143.3 & 52.4 & 8.31(+0.08/$-$0.08) & 8.34 & O3N2 \\
SN2016dsb & IIb & GALEXASC J015900.57-322225.2 & 0.0161 & -- & -- &-- & < 8.15 &N2 \\
SN2016gkg & IIb & NGC 613 & 0.0049 & 122.2 & 35.7 & 8.53(+0.18/$-$0.18) & 8.59 & N2 \\
SN2016iyc & IIb & UGC 11924 & 0.0127 & 120.2 & 61.4 & 8.30(+0.05/$-$0.05) & 8.35 & O3N2 \\
SN2017mw & IIb & ESO 316-G7 & 0.0120 & 158.7 & 70.0 & 8.23(+0.02/$-$0.02) & 8.22 & O3N2 \\
SN2018ddr & IIb & UGC 8896 & 0.0146 & 69.2 & 83.7 & 8.44(+0.09/$-$0.09) & 8.48 & O3N2 \\
SN2018gjx & IIb & NGC 865 & 0.0100 & 159.3 & 90.0 & -- & 8.54 & O3N2 \\
SN2019bao & IIb & UGC 5687 & 0.0119 & 111.4 & 80.0 & 8.28(+0.15/$-$0.15) & 8.38 & O3N2 \\
SN2019pqo & IIb & NGC 5980 & 0.0141 & 14.5 & 76.4 & 8.58(+0.02/$-$0.02) & 8.61 & O3N2 \\
SN2020fqv & IIb & NGC 4568 & 0.0075 & -- & -- & -- & 8.57 & O3N2 \\
SN2021bxu & IIb & ESO 478-G6 & 0.0178 & 101.8 & 57.7 & 8.49(+0.03/$-$0.03) & 8.54 & O3N2 \\
ASASSN-15ta & Ib & GALEXASC J202933.17-615703.5 & 0.0150 & 83.5 & 48.9 & 8.28(+0.12/$-$0.11) & 8.27 & O3N2 \\
ASASSN-16ff & Ib & ESO 218-G008 & 0.0087 & 28.4 & 90.0 & -- & 8.14 & O3N2 \\
AT2015dd & Ib & NGC 5483 & 0.0060 & 18.9 & 26.3 & 8.48(+0.02/$-$0.02) & -- & O3N2 \\
Gaia15acs & Ib & PGC 65805 & 0.0200 & 62.8 & 90.0 & -- & 8.52 & O3N2 \\
\makecell{MASTEROT\\J120451.50\\+265946.6}  & Ib & NGC 4080 & 0.0019 & 121.1 & 75.6 & 8.47(+0.10/$-$0.10) & 8.48 & O3N2 \\
PS15cer & Ib & NGC 7349 & 0.0150 & 165.2 & 76.3 & 8.43(+0.07/$-$0.07) & 8.43 & O3N2 \\
PTF09dfk & Ib & Anonymous & 0.0160 & 99.4 & 44.6 & 8.32(+0.07/$-$0.07) & 8.27 & O3N2 \\
SN2004cc & Ib & NGC 4568 & 0.0075 & 28.6 & 67.5 & 8.50(+0.05/$-$0.05) & 8.55 & O3N2 \\
SN2004dk & Ib & NGC 6118 & 0.0052 & 58.1 & 68.7 & 8.56(+0.02/$-$0.02) & 8.50 & O3N2 \\
SN2006lc & Ib & SDSS J24424.36-000943.4 & 0.0161 & 66.1 & 51.8 & 8.56(+0.04/$-$0.04) & 8.56 & O3N2 \\
SN2009iu & Ib & NGC 7329 & 0.0108 & 107.3 & 42.7 & 8.58(+0.12/$-$0.12) & -- & O3N2 \\
SN2012au & Ib & NGC 4790 & 0.0045 & 87.0 & 58.8 & 8.48(+0.01/$-$0.01) & 8.46 & O3N2 \\
SN2014ge & Ib & NGC 4080 & 0.0019 & 121.1 & 75.6 & 8.43(+0.07/$-$0.07) & 8.48 & O3N2 \\
SN2016ajo & Ib & UGC 11344 & 0.0160 & 162.8 & 64.9 & 8.32(+0.04/$-$0.04) & 8.28 & O3N2 \\
SN2016cdd & Ib & ESO 218-G008 & 0.0087 & 28.4 & 90.0 & -- & 8.12 & O3N2 \\
SN2017ewx & Ib & NGC 5418 & 0.0160 & 45.4 & 68.5 & 8.50(+0.06/$-$0.06) & -- & O3N2 \\
SN2019ehk & Ib & NGC 4321 & 0.0043 & 153.0 & 24.0 & 8.58(+0.03/$-$0.03) & 8.46 & O3N2 \\
SN2019yvr & Ib & NGC 4666 & 0.0050 & 40.6 & 69.6 & 8.58(+0.01/$-$0.01) & 8.48 & O3N2 \\
SN2020admc & Ib & ESO 320-G31 & 0.0100 & 144.7 & 90.0 & -- & 8.53 & O3N2 \\
SN2020hvp & Ib & NGC 6118 & 0.0052 & 58.1 & 68.7 & 8.55(+0.08/$-$0.08) & 8.56 & O3N2 \\
SN2021kos & Ib & IC 719 & 0.0061 & 52.4 & 90.0 & -- & 8.54 & O3N2 \\
SN2023crx & Ib & NGC1602 & 0.0120 & 22.9 & 81.2 & 8.54(+0.07/$-$0.07) & 8.45 & O3N2 \\
iPTF13bvn & Ib & NGC 5806 & 0.0045 & 171.8 & 60.4 & 8.52(+0.04/$-$0.03) & 8.59 & O3N2 \\
ASASSN-15kj & Ic & ESO 297-G37 & 0.0185 & 63.4 & 90.0 & -- & 8.48 & O3N2 \\
ASASSN-21vr & Ic & NGC 3256 & 0.0094 & 87.2 & 48.2 & 8.53(+0.00/$-$0.00) & 8.52 & O3N2 \\
SN2002J & Ic & NGC 3464 & 0.0125 & 110.8 & 50.8 & 8.52(+0.06/$-$0.06) & 8.55 & O3N2 \\
SN2002ao & Ic & UGC 9299 & 0.0051 & 29.8 & 24.7 & 8.26(+0.09/$-$0.08) & 8.37 & O3N2 \\
SN2005lr & Ic & ESO 492-G2 & 0.0086 & 153.6 & 48.8 & 8.49(+0.10/$-$0.10) & -- & O3N2 \\
SN2007rz & Ic & NGC 1590 & 0.0130 & 110.0 & 27.9 & 8.61(+0.03/$-$0.03) & 8.62 & O3N2 \\
SN2009dt & Ic & IC 5169 & 0.0104 & 24.1 & 84.0 & 8.58(+0.04/$-$0.04) & 8.57 & O3N2 \\
SN2010do & Ic & NGC 5374 & 0.0146 & 45.0 & 36.9 & 8.53(+0.02/$-$0.02) & 8.58 & O3N2 \\
SN2011N & Ic & ESO 120-G16 & 0.0114 & 0.6 & 77.4 & 8.50(+0.05/$-$0.05) & 8.53 & O3N2 \\
SN2011jm & Ic & NGC 4809 & 0.0031 & 65.0 & 90.0 & -- & 8.14 & O3N2 \\
SN2013dk & Ic & NGC 4038 & 0.0055 & 160.4 & 51.9 & 8.54(+0.00/$-$0.00) & 8.57 & O3N2 \\
SN2014L & Ic & NGC 4254 & 0.0080 & 60.0 & 20.1 & 8.60(+0.01/$-$0.01) & 8.61 & O3N2 \\
SN2014eh & Ic & NGC 6907 & 0.0106 & 57.7 & 37.5 & 8.50(+0.05/$-$0.05) & 8.41 & O3N2 \\
SN2016iae & Ic & NGC 1532 & 0.0040 & 34.2 & 83.0 & 8.53(+0.07/$-$0.07) & 8.53 & O3N2 \\
SN2017fwm & Ic & ESO 141-IG32 & 0.0160 & 178.8 & 41.9 & 8.53(+0.06/$-$0.06) & -- & O3N2 \\
SN2017rt & Ic & NGC 3836 & 0.0120 & 137.7 & 39.8 & 8.40(+0.02/$-$0.02) & 8.40 & O3N2 \\
SN2019yz & Ic & UGC 9977 & 0.0064 & 79.5 & 90.0 & -- & 8.33 & O3N2 \\
SN2020oi & Ic & MESSIER 100 & 0.0052 & 153.0 & 24.0 & 8.59(+0.01/$-$0.01) & 8.59 & O3N2 \\
SN2021aexi & Ic & NGC 7771 & 0.0140 & 68.0 & 66.7 & 8.58(+0.04/$-$0.04) & 8.56 & O3N2 \\
SN2021ocs & Ic & NGC 7828 & 0.0191 & 136.7 & 90.0 & -- & 8.49 & O3N2 \\
SN2023bqj & Ic & ESO-163-G011 & 0.0090 & 3.3 & 70.9 & 8.48(+0.18/$-$0.18) & 8.49 & O3N2 \\
SN2023cj & Ic & NGC5468 & 0.0095 & 109.2 & 21.1 & 8.31(+0.05/$-$0.05) & -- & O3N2 \\

    \end{longtable}
\end{center}


\bsp	
\label{lastpage}
\end{document}